\documentclass{jpp}
\usepackage{graphicx}

\usepackage[utf8]{inputenc}
\usepackage[T1]{fontenc}
\usepackage{amsmath}
\usepackage{xcolor}

\shorttitle{Compton scattering in particle-in-cell codes}
\shortauthor{F. Del Gaudio, T. Grismayer, R. A. Fonseca and L. O. Silva}

\title{Compton scattering in particle-in-cell codes}

\author{F. Del Gaudio\aff{1}
  \corresp{\email{fabrizio.gaudio@tecnico.ulisboa.pt}},
  T. Grismayer\aff{1},
  R. A. Fonseca\aff{1,2}
 \and L. O. Silva\aff{1}}

\affiliation{\aff{1}GoLP/Instituto de Plasmas e Fus\~ao Nuclear, Instituto Superior T\'ecnico, Universidade de Lisboa, 1049-001 Lisbon, Portugal
\aff{2}DCTI/ISCTE Instituto Universit\'ario de Lisboa, 1649-026 Lisboa, Portugal}

\begin{document}

\maketitle

\begin{abstract}

We present a Monte Carlo collisional scheme that models single Compton scattering between leptons and photons in particle-in-cell codes.
The numerical implementation of Compton scattering can deal with macro-particles of different weights and conserves momentum and energy in each collision.
Our scheme is validated through two benchmarks for which exact analytical solutions exist: the inverse Compton spectra produced by an electron scattering with an isotropic photon gas and the photon-electron gas equilibrium described by the Kompaneets equation.
It opens new opportunities for numerical investigation of plasma phenomena where a significant population of high energy photons is present in the system.

\end{abstract}

\section{Introduction}

Computer simulations for kinetic plasma processes are of core interest for a variety of scenarios, ranging from astrophysics to laboratory experiments.
%Two different approaches exist to model kinetic effects: Vlasov codes that require a fine discretisation of the plasma phase-space, and particle-in-cell (PIC) codes that provide a numerical solution to the Klimontovich equation with macro-particles.
Particle-in-cell methodology \citep{Evans_1957, Dawson_RMP_1983, Bird1989, Birdsall, Hockney} is one of the most popular and widely used technique, which pioneered the study of collisionless plasmas.
The standard PIC loop can be enriched with various Quantum Electrodynamics (QED) cross-sections to investigate astrophysical environments and model laboratory experiments where quantum processes affect the plasma dynamics.
These modules rely on Monte Carlo techniques by taking advantage of the inherent stochasticity of QED processes.
The coupling of QED Monte Carlo modules to the PIC loop represents a unique numerical tool that allows studying such scenarios from first principles.
(Ultra relativistic particles loose their energy via various radiative energy loss channels.) For example, the inclusion of nonlinear Compton scattering (QED synchrotron) is essential to simulate the interaction of matter with ultra-intense electromagnetic fields\citep{Nerush_PRL_2011,Ridgers_PRL_2012,Vranic_PRL_2014,Blackburn_PRL_2014,Gonoskov_PRE_2015,Vranic_NJP_2016,Lobet_JPC_2016,Grismayer_PoP_2016,Vranic_PPCF_2016,Jirka_PRE_2016,Grismayer_PRE_2017}. Several other radiative energy loss channels can participate in the production of high energy photons, such as  curvature radiation, inverse Compton emission, and Bremsstrahlung.
These photons are produced in astronomical sources such as Active Galactic Nuclei, X-ray binaries, supernova remnants, pulsars and gamma-ray bursts can further interact with matter and in particular with the surrounding plasma. The wavelengths of high energy photons are typically smaller than the average inter-particle distance of any tenuous plasma, implying only binary interaction between single photons and electrons. The leading photon - electron (positron) interaction mechanism is single Compton scattering \citep{Compton_PR_M1923}.

The collision of high energy photons with the plasma electrons is at the core of some fundamental scenarios: explains the saturation properties of cyclotron radiation masers \citep{Dreicer_PoF_1964}, the relaxation to the thermal equilibrium of a photon-electron gas \citep{Kompaneets_JETP_1957, Peyraud_JP_1968_I, Peyraud_JP_1968_II, Peyraud_JP_1968_III}, or the Comptonisation of the microwave background \citep{Sunyaev_ARAA_1980}.
These seminal studies approximate the plasma as a gas of free electrons and thus neglect its collective behaviour.
Frederiksen \citep{Frederiksen_APJL_2008} and more recently the Authors \citep{DelGaudio_PRL_2020} have shown that bursts of hard X-rays can couple to the collective plasma dynamics via incoherent Compton scattering events and drive plasma wakes.
Such phenomena can be studied numerically by coupling a Monte Carlo Compton module to the PIC loop \citep{Haugboelle_2005, Haugboelle_PoP_2013}, in a binary collision module.

The implementation of binary collisions in PIC codes is extensively discussed in the literature, with main focus on Coulomb collisions \citep{Takizuka_JCP_1977,Wilson_JGR_1992,Miller_GRL_1994,Vahedi_CPC_1995,Nanbu_PRE_1997,Larson_JCP_2003,Kawamura_PRE_2005,Sherlock_JCP_2008,Sentoku_JCP_2008,Peano_PRE_2009,Turrel_JCP_2015,Higginson_JCP_2017}.
The usual implementation relies on the approximation of small cumulative scattering angles \citep{Takizuka_JCP_1977, Miller_GRL_1994, Nanbu_PRE_1997}, which allows relaxing the simulation time step that is not bound to resolve the collision frequency.
This method improves considerably the computational performance but neglects the effect of large-angle collisions.
Recently, Turrel \citep{Turrel_JCP_2015} and Higginson \citep{Higginson_JCP_2017} included the effect of large angle collisions.
The definition of a cut-off angle allows identifying the occurrence of small-angle collisions or large angle ones, based on the impact parameter of the colliding particles.
In the case of Compton scattering, the method of small cumulative collisions is unworkable, as the angle of scattering ranges within $\theta\in [0,~\pi]$ with similar probability for all angles in the Thomson regime.
In fact, in the Compton regime, the most likely interaction occurs for large scattering angles, and the definition of an impact parameter for Compton scattering is meaningless.
Compton scattering theory considers only initial and final asymptotic electron and photon states neglecting the extent of the interaction in configuration space. As a result, the implementation of a cumulative scattering algorithm does not seem to be applicable.

In this article, we describe the implementation of a single Compton scattering collision module for particle-in-cell codes. It relies on first principles (the Klein-Nishina \citep{Klein_Nat_1928} cross-section is employed with no approximations) and allows a self-consistent treatment of the high-frequency radiation coupling with the plasma dynamics.
In section \ref{sec: Compton}, we review the basic theory for Compton scattering with particular attention to the Lorentz invariant quantities that the model must enforce for reproducing the correct scattering rates in the collision at relativistic energies \citep{Peano_PRE_2009}. 
Section \ref{sec: algorithm} is devoted to the implementation of our collision procedure.
In Section \ref{sec: benchmark}, we benchmark our code against problems for which exact analytical solution or formulation exist, namely the scattering photon spectrum of a relativistic charge \citep{Blumenthal_RMP_1970}, and the Kompaneets equation \citep{Kompaneets_JETP_1957}.
Finally, in Section \ref{sec: performance}, we comment on the computational cost that our module brings as compared to a standard PIC loop. Summary and conclusions are presented in Section \ref{sec: summary}.

%----------------------------------------------------------------------------------------

\section{Compton scattering} \label{sec: Compton}

Single Compton scattering is the inelastic collision between a photon and an electron \citep{Compton_PR_M1923}.
It is the generalization of Thomson scattering \citep{Thomson_CETG_1906}, for any value of the incident photon of energy $\hbar\omega$ in the electron proper frame of reference.
By applying energy and momentum conservation in the electron rest frame, later denoted reference frame (O),
\begin{eqnarray} 
\hbar\omega + mc^2 & = & \hbar\omega' + \gamma'mc^2, \\
\hbar{\bf k} & = & \hbar{\bf k}'+{\bf p}', 
\end{eqnarray}
where $\gamma'=\sqrt{1+p'^2/m^2c^2}$, and $\omega=ck$, the photon frequency shift over one collision is
\begin{equation}
\frac{\omega'}{\omega} = \frac{mc^2}{mc^2+\hbar\omega(1-\cos\theta)}, \label{eq: frequency shift}
\end{equation}
where $\omega$ ($\omega'$) is the absorbed (emitted) frequency, and $\theta$ is the scattering angle.
For $\hbar\omega\ll mc^2$, the Thomson limit $\omega'\simeq\omega$ is recovered.
However, when the incident photon energy approaches and exceeds the electron rest mass energy $\hbar\omega\gtrsim mc^2$, the energy transfer becomes relevant.
For $\hbar\omega\gg mc^2$, at $\theta\simeq -\pi$, the photon transfers to the electron up to half its energy $\hbar(\omega-\omega')\simeq \hbar\omega/2$. 
The classical theory of radiation explains Thomson scattering in terms of plane wave absorption and consequent dipole radiation from the oscillating charge~\citep{Landau, Jackson}, but does not predict Compton scattering, which is intrinsically a quantum process.

\subsection{Klein-Nishina cross section} %----------------------------------------------------------------------------------------

In the rest frame of an electron, the single Compton scattering probability is determined by the Klein-Nishina differential (in solid angle $\Omega$) cross section~\citep{Klein_Nat_1928}, which, for unpolarised photons, reads
\begin{equation}
\frac{d\sigma}{d\Omega} = \frac{r_e^2}{2}\left(\frac{\omega'}{\omega}\right)^2\left(\frac{\omega'}{\omega}+\frac{\omega}{\omega'}-\sin^2\theta\right) \label{eq. KN}
\end{equation}
where $r_e=e^2/mc^2$ is the classical electron radius.
By combining Eq.(\ref{eq: frequency shift}) and Eq.(\ref{eq. KN}), and integrating over the solid angle $d\Omega=\sin\theta d\theta d\phi$ ($\phi$ is the symmetry angle around the direction of the incoming photon) the total cross section reads
\begin{equation}
\sigma(\epsilon)= \frac{\pi r_e^2}{\epsilon}\left[\left(1-\frac{2}{\epsilon}-\frac{2}{\epsilon^2}\right)\log(1+2\epsilon)+\frac{1}{2}+\frac{4}{\epsilon}-\frac{1}{2(1+2\epsilon)^2} \right]
\end{equation}
where $\epsilon=\hbar\omega/mc^2$.
In the limit for low photon energies $$\lim_{ \epsilon \to 0} \sigma(\epsilon)=\sigma_T$$ the Thomson cross section is recovered.
For high photon energies $\epsilon\gg 1$ the cross section has the limiting expression $$\lim_{ \epsilon \gg 1} \sigma(\epsilon)=\frac{3}{8}\sigma_T\frac{\log(2\epsilon)}{\epsilon}$$ and decreases with respect to the incident photon energy.

\subsection{Relativistic kinematics and Lorentz invariants} %----------------------------------------------------------------------------------------

\begin{figure}\centering
\vspace{0.5cm}
  \centerline{\includegraphics[width=0.8\linewidth]{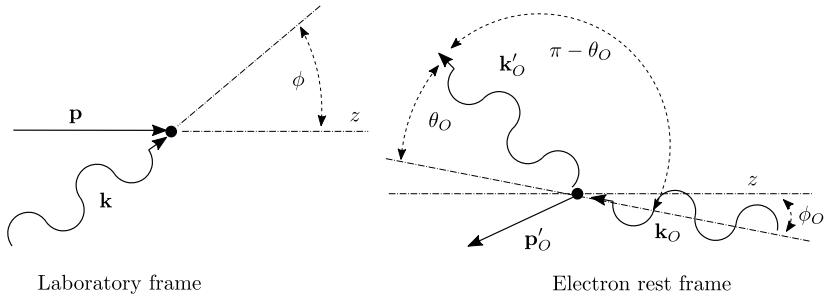}}
  \caption{Schematic of the Compton scattering relativistic kinematics. }
\label{fig: kinematic}
\end{figure}

We consider a relativistic electron which propagates along the z coordinate at velocity $\beta c$ and scatters with a photon at an incident angle $\phi$ in the laboratory frame, see Fig. \ref{fig: kinematic}.
In the electron proper frame of reference $_O$, the incident angle is modified by relativistic effects.
In the frame $_O$ the incident photon is confined within a small cone \citep{Blumenthal_RMP_1970}
\begin{equation}
\tan\phi_O=\frac{\sin\phi}{\gamma(\cos\phi-\beta)} \label{eq: cmp cone}
\end{equation}
of aperture $1/\gamma$.
The photon energy in the frame $_O$ reads
\begin{equation}
\epsilon_O=\gamma\epsilon(1-\beta\cos\phi). \label{eq: ene to O}
\end{equation}
It varies in the range $\epsilon_O\in[~\epsilon/2\gamma,~2\gamma\epsilon~]$ according to the incident angle $\phi$.
In the $_O$ frame, the photon energy after scattering obeys Eq. (\ref{eq: frequency shift}) and in the laboratory frame reads
\begin{equation}
\epsilon'=\gamma\epsilon_O'[1+\beta\cos(\pi-\theta_O-\phi_O)]\simeq \gamma\epsilon_O'(1-\cos\theta_O ), \label{eq: ene from O}
\end{equation}
due to the Lorentz transformation, where $\beta\simeq1$ and $\phi_O\sim1/\gamma$.
In the Thomson regime, $\omega_O'\simeq\omega_O$ and the maximum energy achieved over one collision is $\epsilon'\simeq4\gamma^2\epsilon$, for $\phi\simeq\pi$ and $\theta_O\simeq\pi$.
In the extreme Klein-Nishina limit, the maximum energy achieved over one collision can be obtained by combining Eqs. (\ref{eq: frequency shift}), (\ref{eq: ene to O}), and (\ref{eq: ene from O}), and reads $\epsilon'\simeq\gamma$.

We now consider the scattering between photons, with distribution function $f_{\omega}$, and electrons, with distribution function $f_e$.
Within a portion of space-time $d{\bf x}dt$, the number of collisions is a Lorentz invariant quantity \citep{Groot} that is given by
\begin{equation}
N = \sigma({\bf p},{\bf k}) c f_{\omega}d{\bf k} f_{e}d{\bf p} d{\bf x}dt.
\end{equation}
In general, the cross section $\sigma({\bf p},{\bf k}) $ depends on the electron momentum ${\bf p}$, and on the photon wavevector ${\bf k}$.
As the space-time element $d{\bf x}dt$, the distribution functions $f_{\omega}$ and $f_{e}$, and the speed of light $c$ are Lorentz invariant therefore $\sigma({\bf p},{\bf k})d{\bf k}d{\bf p}$ is also Lorentz invariant~\citep{Landau}.
This invariance allows to obtain the cross section in any inertial frame ($\gamma=\sqrt{1+{\bf p}^2/m^2c^4}$, $\epsilon=\hbar|{\bf k}|/mc$).
Knowing the cross section in the electron proper frame of reference ($\gamma_O=1$, $\epsilon_O=\gamma\epsilon-\hbar{\bf p}\cdot{\bf k}/m^2c^2$) 
\begin{equation}
\sigma({\bf p},{\bf k})d{\bf k}d{\bf p} =  \sigma(\omega_O)d{\bf k_O}d{\bf p_O},
\end{equation}
we finally obtain
\begin{equation}
\sigma({\bf p},{\bf k}) =  \sigma(\epsilon_O)\frac{\epsilon_O}{\gamma\epsilon}, \label{eq. sigmaINV}
\end{equation}
since $d{\bf k}/\epsilon$ and $d{\bf p}/\gamma$ are Lorentz invariants~\citep{Landau}.
%----------------------------------------------------------------------------------------

\section{Single Compton scattering algorithm} \label{sec: algorithm}

\begin{figure}\centering
\vspace{0.5cm}
  \centerline{\includegraphics[width=0.8\linewidth]{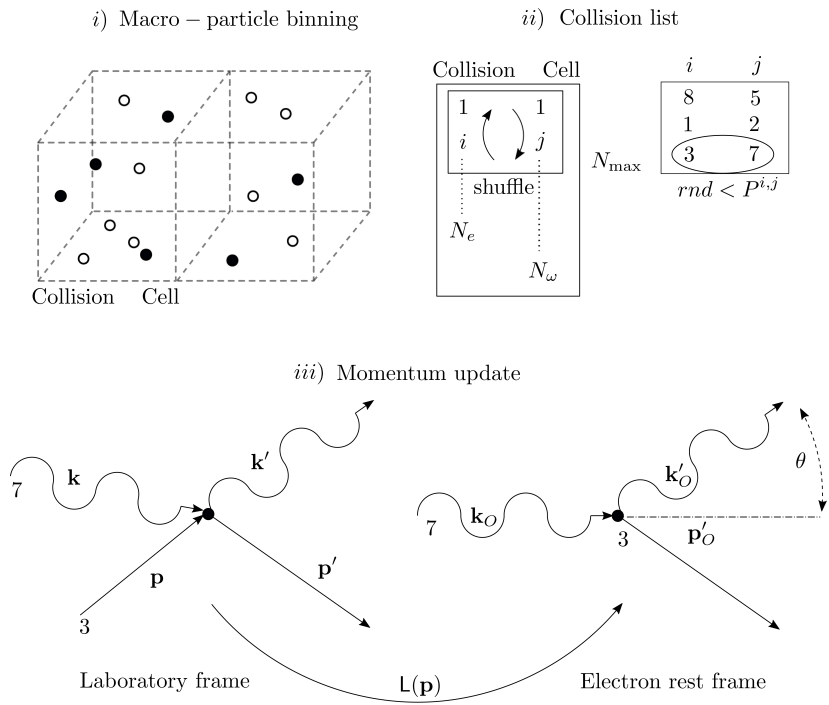}}
  \caption{Schematic of the Compton scattering algorithm. It follows three steps: i) the macro-particles are binned into collision cells $\Delta{\bf x}$, ii) the probability $P^{i,j}$ of interaction within $\Delta{\bf x}\Delta t$ is computed and scattering macro-particles are chosen using the no-time-counter method, iii) the momenta of the scattering macro-particles are updated.}
\label{fig. algo}
\end{figure}

The implementation of single Compton scattering in a PIC code must not only recover the correct microphysics of the process (frequency shift, angle, momentum recoil) but must preserve the invariant number of collisions to obtain the correct scattering rates \citep{Peano_PRE_2009}.
The implementation follows naturally as each macro-particle represents an ensemble of real particles that are close neighbors in phase space.
Each macro-particle has a weight $q$ that relates to the number of real particles it represents, and thus samples a portion of the distribution function of real particles.
Figure~\ref{fig. algo} outlines our implementation that follows three steps: i) binning of the macro-particles into collision cells $\Delta{\bf x}$, a volume in configuration space, ii) pairing of the colliding macro-particles according to their probability $P^{i,j}$ of interaction within $\Delta{\bf x}\Delta t$, iii) update of the momenta of the scattering macro-particles.

\subsection{Macro-particles binning} %----------------------------------------------------------------------------------------

The binning of macro-particles in collision cells naturally uses the single PIC cell as the smallest binning volume.
The size of a PIC cell is also the smallest scale over which the self-consistent plasma collective fields are computed.
For this reason, the collision cells are usually set equal to the PIC cells.
Macro-photons and macro-electrons are binned in the collision cells and sorted such that we identify the indexes of macro-electrons and macro-photons within each collision cell.

\subsection{Pairing} %----------------------------------------------------------------------------------------

For each collision cell, we pair the scattering couples and add them to a scattering list using the no-time-counter method (NTC) \citep{Bird1989, Abe_CF_1993}.
The NTC method is a popular Monte Carlo scheme for collision procedures involving single scattering events (not for cumulative scattering).
The standard pairing routines for cumulative Coulomb collisions allow for a time step larger than the collision frequency, thus all macro-particles are involved in the scattering process each time step.
Instead, the NTC method applies when the time step resolves the collision frequency such that the maximum possible number of macro-scatterings within a time step involves only a subset of all the macro-particles.
Developed three decades ago~\citep{Bird1989}, NTC provides a cost reduction for the sampling of a discrete probability distribution function.
We detail now the NTC algorithm applied to single Compton scattering.
\\
\\
We consider a collision cell containing $N_{\omega}$ macro-photons and $N_e$ macro-electrons.
A conservative upper-bound to the maximum probability of any macro-particle to collide within $\Delta t$ is 
\begin{equation}
P_{\mathrm{max}} = 2 \sigma_T c \Delta t ~\mathrm{max}[q^i_e,q^j_{\omega}]
\end{equation}
where $\mathrm{max}[q^i_e,q^j_{\omega}]$ is the largest weight with units of a density among all macro-particles in the collision cell ($i\in[1,N_e]$ macro-electrons and $j\in[1,N_{\omega}]$ macro-photons).
The factor $2$ appears conservatively as the upper bound in the relativistic transformation of the cross section $\sigma = \sigma_{T,O}\epsilon_O/\gamma \epsilon$, where $\mathrm{max}(\epsilon_O)=2\gamma\epsilon$.
The maximum number of macro-particles that can scatter $N_{\mathrm{max}}$ is given by the maximum probability $P_{\mathrm{max}}$ times the number of all the possible unsorted pairing combinations $N_eN_{\omega}$ (potential scatterings) of the macro-photons with the macro-electrons.
It reads
\begin{equation}
N_{\mathrm{max}}=P_{\mathrm{max}} N_e N_{\omega}.
\end{equation}
The number $N_{\mathrm{max}}$ is usually not an integer and is rounded to the next or previous integer by a Monte Carlo sampling of the residue.
This procedure preserves statistically the correct number of collisions within $\Delta{\bf x}\Delta t$.
We randomly pair $N_{\mathrm{max}}$ macro-photons and $N_{\mathrm{max}}$ macro-electrons.
This follows two steps: i) the random sorting of the macro-photons and the macro-electrons, ii)  the selection of the first $N_{\mathrm{max}}$ indexes.
At this point, we have a shortlist of $N_{\mathrm{max}}$ randomly paired macro-particles, which contains the maximum possible scatterings in the collision cell.
For each pair in the short list, a random number $rnd\in[0,1]$ is rolled and compared with the joint probability
\begin{equation}
P^{i,j}=\sigma({\bf p}^i,{\bf k}^j) c \Delta t ~\mathrm{max}[q^i_e,q^j_{\omega}]/P_{\mathrm{max}}
\end{equation}
of scattering after having being selected within the $N_{\mathrm{max}}$ pairs.

To compute the cross section $\sigma({\bf p}^i,{\bf k}^j)$ we proceed as follows.
The energy of the photon ${\bf k}^j$ is Lorentz boosted in the rest frame of the electron ${\bf p}^i$
\begin{equation}
\epsilon_O^j=\gamma^i\epsilon^j-\hbar{\bf p}^i\cdot {\bf k}^j/m^2c^2.
\end{equation}
Then, the cross-section $\sigma_C(\epsilon_O^j)$ is computed in this frame and boosted back into the simulation frame using eq.~\ref{eq. sigmaINV}.
A macro-electron/macro-photon pair from the shortlist is admitted to the scattering list based on a rejection method (accepted if $rnd<P^{i,j}$).

\subsection{Momentum update} %----------------------------------------------------------------------------------------

For each pair in the scattering list, the momenta are updated according to the Compton frequency shift and momentum recoil.
The macro-photon four-wavevector $\mathcal{K}=(\epsilon, \hbar{\bf k}/mc)$ is Lorentz boosted in the rest frame of the electron, of momentum ${\bf p}$ in the simulation frame, as $\mathcal{K}_O=\mathbb{ L}({\bf p})\mathcal{K}$, where the boost matrix is
\begin{equation}
\setlength{\arraycolsep}{0pt}
\renewcommand{\arraystretch}{1.3}
\mathbb{ L}({\bf p}) = \left[
\begin{array}{cc}
  \gamma  &  -{\bf p}/mc   \\
  \displaystyle
     -{\bf p}^T/mc &\quad \mathbb{I}+{\bf p}^T{\bf p}/m^2c^2(1+\gamma)  \\
\end{array}  \right] ,
\label{eq. boost}
\end{equation}
and $\mathbb{I}$ the $3\times3$ identity matrix.
In the frame $O$, we identify the unit vector along the photon propagation direction $\hat{\bf k}_0$ which defines the symmetry axis for the scattering.
We define an orthonormal unit vector base $\hat{\bf e}_1 = \hat{\bf k}_0, \hat{\bf e}_2\perp \hat{\bf e}_1, \hat{\bf e}_3=\hat{\bf e}_1\times \hat{\bf e}_2$.
The two scattering angles $\theta$ and $\phi$ are then sampled, $\theta$ is the angle with respect to $\hat{\bf e}_1$, and $\phi$ is the angle on the plane $\hat{\bf e}_2,\hat{\bf e}_3$.
This latter, being the angle of rotational symmetry, is chosen randomly between $0$ and $2\pi$.
The angle $\theta$, or rather the parameter $\mu=\cos\theta$ is obtained by the Inverse Transform Sampling method of the cumulative probability function given by the differential cross-section of the process (see Appendix \ref{App: ITS}).
We preferred this method rather than a rejection method, whose efficiency decreases for $\epsilon_O\gtrsim 1$ due to the steepening of the probability density function close to $\mu\simeq -1$. 
The scattered photon energy is $\epsilon_O'$ given by eq.~\ref{eq: frequency shift}
\begin{equation}
\epsilon_O' = \frac{\epsilon_O}{1+\epsilon_O(1-\mu)}
\end{equation}
and the scattered wavevector is
\begin{equation}
\frac{\hbar{\bf k}_O'}{mc} = \epsilon_O'\left(\mu\hat{\bf e}_1+\sqrt{1-\mu^2}\cos\phi\hat{\bf e}_2 +\sqrt{1-\mu^2}\sin\phi\hat{\bf e}_3\right)
\end{equation}
We transform back to the simulation frame $\mathcal{K}_O'=(\epsilon_O',\hbar{\bf k}_O'/mc)$ simply as $\mathcal{K}'=\mathbb{ L}(-{\bf p})\mathcal{K}_O'$, and by conservation of momentum the scattered electron has a new momentum ${\bf p}'={\bf p}+\hbar({\bf k}-{\bf k}')$.

\subsection{Macro-particles with difference in weight} %----------------------------------------------------------------------------------------

In PIC codes, it is unlikely that two scattering particles possess the same weight.
Two main techniques to approach the problem have been discussed in \citep{Sentoku_JCP_2008}.
The first approach is based on a rejection method for which the scattering occurs with a probability based on the weights of the two-scattering macro-particles.
This method does not reproduce the energy and momentum transfer of each collision but only on average, for a sufficiently high number of macro-particle in the collision cell.
To preserve the energy and momentum transfer per collision, an alternative is first to spilt the scattering macro-particle of weight $q$ into a scattering fraction $q_s$ (${\bf p}_s$) and a non-scattering fraction $q_{ns}$ (${\bf p}$).
After the collision takes place, the two fractions are merged again into the macro-particle $q$, which has the average energy and momentum of $q_s$ and $q_{ns}$.
This last method becomes inaccurate when ${\bf p}_s$ differs significantly from ${\bf p}$ such that the two fractions $q_s$ and $q_{ns}$ refer to two well distinct portions of the phase space.
This issue has already been addressed by \citep{Haugboelle_2005} and relies on splitting and merging at two different steps.
Here, we address this problem similarly but only the largest weight macro-particle is split before scattering. We briefly recall the main steps of the splitting procedure:
\begin{itemize}
    \item  Identify the two scattering macro-particles of weight $q_e^i$ and $q_{\omega}^j$, and select the largest weight between the two: $\mathrm{max}[q_e^i,q_{\omega}^j]$,
    \item  Create a new particle of weight equal to $\mathrm{min}[q_e^i,q_{\omega}^j]$,
    \item  The two macro-particles of equal weight $\mathrm{min}[q_e^i,q_{\omega}^j]$ are now paired and can Compton scattered as described before,
    \item  reassign to the split macro-particle the weight $\mathrm{max}[q_e^i,q_{\omega}^j]-\mathrm{min}[q_e^i,q_{\omega}^j]$.
\end{itemize}
The splitting is performed within the scattering routine and can lead to a significant increase of macro-particles in the simulation.
Merging algorithms \citep{Vranic_CPC_2015} can be used at a different step of the PIC loop to preserve the number of macro-particles in the simulation within a reasonable maximum, thus avoiding their exponential increase.
The advantage of merging at a later stage is that only macro-particles close in phase space merge.
Details can be found in \citep{Vranic_CPC_2015}.

%----------------------------------------------------------------------------------------

\section{Benchmarks} \label{sec: benchmark}

To benchmark our algorithm, we choose two problems that possess an exact analytical solution:
\begin{itemize}
\item the inverse Compton spectra produced by an electron scattering with an isotropic photon gas \citep{Blumenthal_RMP_1970},
\item the relaxation to the thermal equilibrium of a photon gas by Compton collisions with a thermal electron gas of fixed non-relativistic temperature described by the Kompaneets equation~\citep{Kompaneets_JETP_1957}.
\end{itemize}

\subsection{Inverse Compton spectra} %----------------------------------------------------------------------------------------

\begin{figure}\centering
  \centerline{\includegraphics[width=0.8\linewidth]{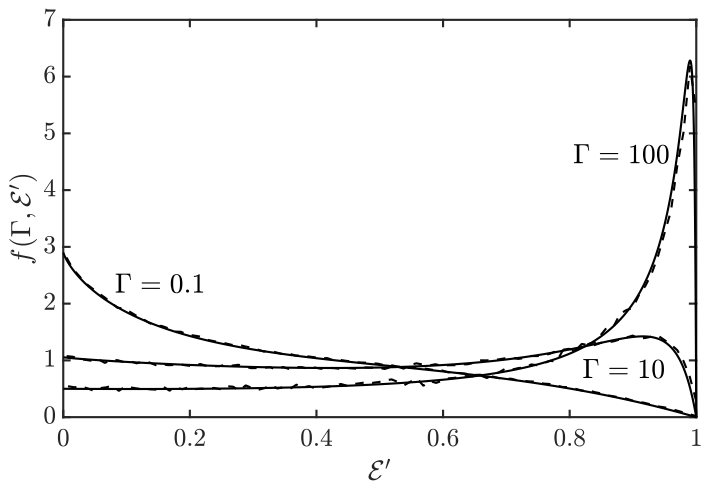}}
  \caption{Scattered photon distribution function $f(\Gamma,\mathcal{E}')$ for $\Gamma=0.1,~10,~100$. The simulation results are shown by dashed lines and the theory, Eq. (\ref{eq: BlumSpec}), by solid lines\citep{Blumenthal_RMP_1970}.}
\label{fig: SBench}
\end{figure}

Blumenthal and Gould \citep{Blumenthal_RMP_1970} derived the inverse Compton spectra produced by the collision of a relativistic electron, $\gamma\gg 1$, with an isotropic gas of photons (see Appendix \ref{App: ICS}).
The scattered photon distribution function reads
\begin{equation} \label{eq: BlumSpec}
f(\Gamma,\mathcal{E}') = 2q\log q +(1+2q)(1-q)+\frac{1}{2}\frac{\Gamma^2 q^2}{1+\Gamma q}(1-q)
\end{equation}
where $q=\mathcal{E}'/[1+\Gamma(1-\mathcal{E}')]$, $\mathcal{E}'=\epsilon'/\epsilon'_{\mathrm{max}}$ is the scattered photon energy normalised to its maximum $\epsilon'_{\mathrm{max}}=\gamma\Gamma/(1+\Gamma)$.
The parameter $\Gamma=4\epsilon\gamma$ relates to the energy of the scattering photons in the electron rest frame and distinguishes two regimes: i) Thomson limit $\Gamma\ll1$ and ii) extreme Klein-Nishina limit $\Gamma\gg1$.

Figure~\ref{fig: SBench} shows the excellent agreement between our simulations (dashed lines) and theory (solid lines), Eq. (\ref{eq: BlumSpec}) for $\Gamma=0.1,~10,~100$.
The scattered photon distribution function $f(\Gamma,\mathcal{E}')$ is normalised $\int d\mathcal{E}' f(\Gamma,\mathcal{E}') =1$.
In our simulations, the photon gas is initialised with $1.5\times10^7$ macro-photons which mimic an emission line.
All macro-photons have the same energy and are propagating in random directions, uniformly distributed on the surface of a sphere in momentum space.
We considered the interaction at different photon energies $\epsilon=0.00025,~0.025,~0.25$.
An equal number of macro-electrons is initialised at a Lorentz factor of $\gamma=100$, all collimated in one direction.
To avoid that each macro-photon scatters more than once the simulation runs for only a single time step where about $1\times10^6$ macro-scatterings occur.
The only constraint on $\Delta t$ is to be low enough such that $P_{\mathrm{max}}<1$, to prevent multiple collisions of the photons to occur within a single time step.

\subsection{Photon-electron gas equilibrium (Kompaneets equation)} %----------------------------------------------------------------------------------------

\begin{figure}\centering
  \centerline{\includegraphics[width=0.8\linewidth]{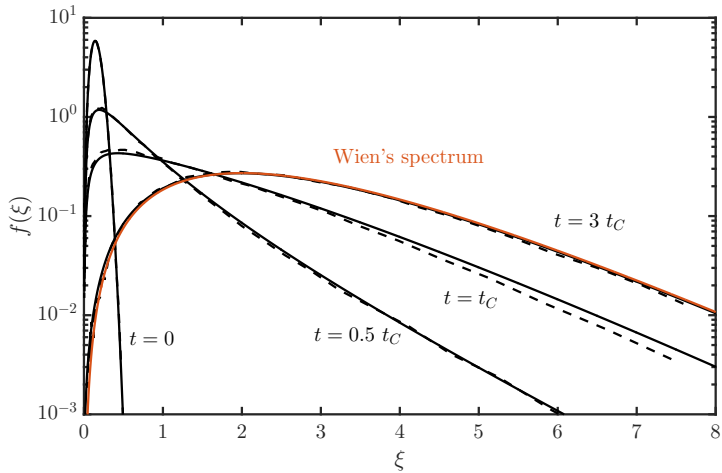}}
  \caption{Time evolution of the photon distribution function $f(\xi)$ by the interaction with an electron gas of density $10^{18}~cm^{-3}$ at $5$ keV temperature for times $t=0,~0.5,~1,~3~t_C$. After $t=3~t_C$ the photon distribution function resembles the Wien spectrum and does not evolve significantly. Simulations in dashed lines and solution of the linear Kompaneets equation, Eq. (\ref{eq: linKomp}), in solid lines~\citep{Kompaneets_JETP_1957}.}
\label{fig: Kompaneets}
\end{figure}
Kompaneets addressed the thermodynamic equilibrium establishing between photons and free electrons if their interaction is only mediated by Compton scattering events \citep{Kompaneets_JETP_1957}.
Kompaneets derived the partial differential equation that describes the temporal evolution of the photon occupation number $n$ resulting from the interaction with an electron gas of fixed nonrelativistic temperature $k_BT\ll mc^2$, where $k_B$ is the Boltzmann constant.
The full collision operator reads
\begin{equation} \label{eq: fullBoltz}
\frac{\partial n}{\partial t} = c\int d{\bf p} \frac{d\sigma}{d\Omega}\left[f_e'n'(1+n)-f_en(1+n')\right]
\end{equation}
where $f_e=f_e(\gamma)$ and $f_e'=f_e(\gamma')$ refer to the electron energy distribution function evaluated at a Compton transition $\gamma mc^2 + \hbar\omega \rightleftharpoons \gamma'mc^2+\hbar\omega'$.
The evaluation of the photon occupation numbers $n=n(\omega)$ and $n'=n(\omega')$ follow the same definition.
The $n^2$ terms account for the photon Bose-Einstein statistics when phenomena like stimulated scattering and superposition of states are considered.
The full Boltzmann operator can be reduced to a Fokker-Planck form within the Thomson limit $\hbar\omega\ll mc^2$ (see Appendix \ref{App: Kompaneets}).
In regimes where the photon occupation number is small, $n\ll1$, the photon electron gas interaction is mediated by single Compton scattering events and the linear Kompaneets equation in terms of the photon energy distribution function $f=\xi^2 n$ reads
\begin{equation} \label{eq: linKomp}
\frac{\partial f}{\partial y} = \frac{\partial }{\partial \xi} \left[\xi^2 \frac{\partial f}{\partial \xi} +\left(\xi^2-2\xi\right) f \right]
\end{equation}
where $y = t/t_C$, $t_C = mc/\sigma_Tn_ek_BT$ is the characteristic relaxation time, $\xi=\hbar\omega/k_BT$ is the photon energy  normalised to the electron temperature. 
\\
\\ 
Figure~\ref{fig: Kompaneets} shows the excellent agreement between our algorithm and the numerical solution of the linear Kompaneets equation, Eq. (\ref{eq: linKomp}), obtained with a finite-difference centred scheme.
In our simulation, more than $10^5$ macro-photons are initialised to mimic an emission line at an average energy of $\bar{\xi}=\langle\xi\rangle=0.2$.
The emission line has a small energy spread of $\sigma_{\xi}^2=\langle \xi^2-\bar{\xi}^2\rangle=0.1$, and the initial distribution
\begin{equation}
f(\xi,~y=0) \propto \exp\left[-\frac{\left(\xi-\bar{\xi}\right)^2}{2\sigma_{\xi}^2}\right]
\end{equation}
is Maxwellian.
The same number of macro-electrons is sampled according to a Maxwellian distribution at a temperature of $k_BT=5$ keV.
To enforce a constant electron temperature during the simulation for a rigorous comparison with theory, we turn off the Lorentz force, which will arise from fluctuations in the electron density.
We also omit the momentum, and energy ceased by the electron to the photons at each collision such that the electron population does not cool down.
At $t\gtrsim3~t_C$, the photon energy distribution reaches equilibrium and converges towards the Wien's spectrum $f\propto\xi^2\exp\left(-\xi\right)$, the correct equilibrium for the linear Kompaneets equation, as expected from the underlying hypothesis.

%----------------------------------------------------------------------------------------

\section{Considerations on the algorithm performance} \label{sec: performance}

\begin{figure}\centering
\centerline{\includegraphics[width=0.8\linewidth]{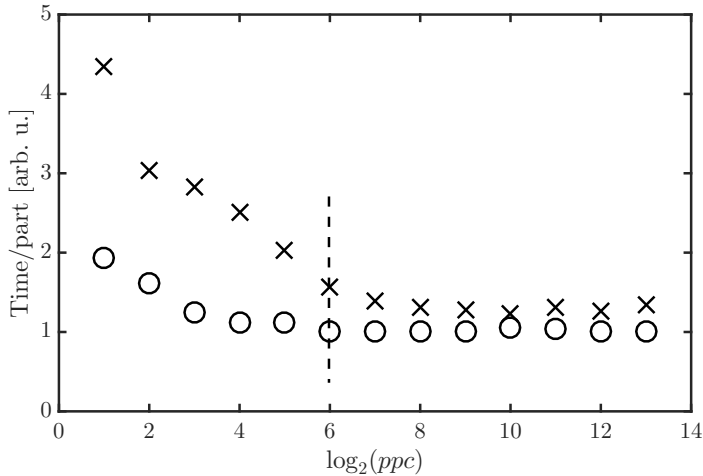}}
\caption{Time of a PIC loop per simulation particle with ($\times$) and without ($\circ$) Compton collisions as a function of the number of particles per cell (ppc). For a low number of ppc, the loop time is determined by the particle sorting routine (dependent also on the number of grid cells) used by the Compton module. For a sufficiently high number of ppc, the scaling of our collision algorithm is proportional to the amount of simulated particles.}
\label{fig: per time}
\end{figure}

In this section, we compare the computational cost of our Compton scattering algorithm with the standard PIC loop.
The computational performance of our algorithm is usually dependent on the physical parameters of the particular simulated system.
A thorough benchmark of its performance should then cover a variety of physical parameters of relevant case scenarios.
In collisional plasmas, a typical benchmark of the performance of a collisional algorithm relies on the simulation of thermal plasma with and without collisions.
Our choice for the comparison follows a similar criterion.
We simulate in 1D a thermal plasma in equilibrium with a photon gas, both at a temperature of $5~\mathrm{keV}$.
The plasma density is $n_p=10^{18}~\mathrm{cm^{-3}}$ and the photon density is $n_{\omega}=3\times 10^{27}~\mathrm{cm^{-3}}$, chosen such that the electron Compton collision frequency is a tenth of the plasma frequency $ c \sigma_T n_{\omega}=\omega_p/10$.
The electrons follow a Maxwell-Boltzmann distribution $f\propto \sqrt{W_k}\exp(-W_k/k_BT)$, and the photons follow a Wien distribution $f\propto W_k^2\exp(-W_k/k_BT)$.
The computational domain is divided into $240$ cells.
The time-step is $\Delta t = 0.099~\omega_p^{-1}$.
Periodic boundary conditions are used.
Figure \ref{fig: per time} shows the time of a PIC loop per simulation particle with ($\times$) and without ($\circ$) Compton collisions as a function of the number of particles per cell (ppc).
For a low number of ppc, the loop time is determined by the particle sorting routine (dependent also on the number of grid cells) used by the Compton module. For a high number of ppc, the scaling of our collision algorithm is proportional to the number of simulated particles. The computational cost of the sorting routine scales with both the number of simulated macro-particles and the number of cells in which they are sorted.
Therefore, there is a trade off, which for the set of parameters of these simulations occurs around $64~ppc$ and is highlighted by the dashed line in Fig. \ref{fig: per time}.
Beyond the region delimited by the dashed line (which changed depending on the grid size and the number of particle per cells), the inclusion of the Compton algorithm does not impact significantly the standard PIC loop performance.
This trade-off must be assessed for the different  numerical parameters/configurations to determine the optimal performance conditions.
%Figure \ref{fig: per slowdown} shows the slowdown of the Compton algorithm as compared to the standard PIC, function of the number of particles per cell (ppc). The slowdown is moderate and the simulations including Compton collisions last from $1.2$ to $2.3$ the time of a standard PIC simulation. After the initial overhead, the scaling of our collision algorithm reduces the slowdown below $40\%$.

%----------------------------------------------------------------------------------------

\section{Summary} \label{sec: summary}

We presented a collision algorithm which incorporates the effect of single Compton scattering from high frequency photons in particle-in-cell codes.
This allows a self-consistent treatment of the high frequency radiation coupling with the plasma dynamics from first principles.
The algorithm shows excellent agreement with respect to the benchmarks: scattering photon spectrum from the collision with relativistic electrons~\citep{Blumenthal_RMP_1970} and the relaxation to thermal equilibrium of a photon population with an electron gas~\citep{Kompaneets_JETP_1957}.
This framework is at the forefront for the numerical modelling of photon-plasma interaction and opens new and exciting opportunities in the numerical investigation of plasma phenomena where a significant population of hard photons is present in the system.
%As a simple example, consider the effect of Coulomb collisions in a plasma. The redistribution of energy and momentum tends to shorten the microscopic currents in the plasma. Different is the case of Compton scattering. As it involves only one charge in the collision, the redistribution of energy and momentum ceased by the photons to the plasma will induce microscopic plasma currents which may become sources of macroscopic fields.

\begin{acknowledgments}
This work was supported by the European Research Council (ERC-2015-AdG Grant 695088), FCT (Portugal) grants SFRH/IF/01780/2013 and PD/BD/114323/2016 in the framework of the Advanced Program in Plasma Science and Engineering (APPLAuSE, FCT grant No. PD/00505/2012).
Simulations were performed at IST cluster (Portugal), and at MareNostrum (Spain) under a PRACE award.
\end{acknowledgments}

% susie put cite commands here, don't bother with citet etc just yet.

\appendix

\section{Scattering angle by the Inverse Transform Sampling method} \label{App: ITS}

The probability distribution function (pdf) of the scattered macro-photon over the scattering angle $\mu=\cos\theta\in[1,-1]$ reads
\begin{equation}
\mathrm{pdf}(\mu, \epsilon_O) = \frac{1}{\sigma(\epsilon_O)}\frac{d\sigma}{d\mu}~, \qquad \mathrm{and}\quad \int_1^{-1} d\mu \frac{d\sigma}{d\mu}=\sigma(\epsilon_O)
\end{equation}
with
\begin{equation}
\frac{d\sigma}{d\mu}=-\pi r_e^2\left(\frac{1}{1+\epsilon_O(1-\mu)}\right)^2\left(\frac{1}{1+\epsilon_O(1-\mu)}+\epsilon_O(1-\mu)+\mu^2\right) 
\end{equation}
The cumulative distribution function (cdf) is
\begin{equation}
\mathrm{cdf}(\mu, \epsilon_O) = \frac{1}{\sigma(\epsilon_O)}\int_1^{\mu}d\mu'\frac{d\sigma}{d\mu'}
\end{equation}
with
\begin{eqnarray}
\int_1^{\mu}d\mu'\frac{d\sigma}{d\mu'} &=& \frac{\pi r_e^2}{\epsilon_O}\left\lbrace \left( 1-\frac{2}{\epsilon_0}-\frac{2}{\epsilon_0^2}\right)\log\left[1+\epsilon_O(1-\mu)\right] \right.\nonumber \\
&& \left. +\frac{1-\mu}{\epsilon_O}\left[1+\frac{1+2\epsilon_O}{1+\epsilon_O(1-\mu)}\right] \right. \nonumber \\
&& \left. +\frac{1}{2} -\frac{1}{[1+\epsilon_O(1-\mu)]^2}     \right\rbrace
\end{eqnarray}
In the inverse transform sampling method a random number is generated in the range $rnd\in[0,1]$, then $\mu=cdf^{-1}(rnd,\epsilon_O)$.
Given the nonlinear dependence of the cdf on $\mu$, we use the bisection method to solve $cdf(\mu,\epsilon_O)-rnd=0$.

\section{Photon spectrum: single scattering with a relativistic electron} \label{App: ICS}

We briefly recall the main steps in the derivation of Eq. (\ref{eq: BlumSpec}), see Ref. \citep{Blumenthal_RMP_1970}.
If the photon gas is isotropic in the laboratory frame, it appears beamed at a small angle $\sim1/\gamma$ in the proper frame $_O$ of reference of an incident relativistic electron $\gamma\gg1$, as shown by Eq. (\ref{eq: cmp cone}).
The Compton scattering differential rate in the laboratory frame reads
\begin{equation}
\frac{dN_{\omega}}{dtd\epsilon'} = \int d\epsilon_O~\int d\Omega_O \frac{dN}{dt_O d\epsilon_Od\Omega_O d\epsilon_O'} \frac{dt_O}{dt}\frac{d\epsilon_O'}{d\epsilon'}. \label{eq: diff rate lab}
\end{equation}
The time interval in the frame $_O$ is $dt_O= dt/\gamma$, and the energy transforms according to Eq. (\ref{eq: ene from O}) as $d\epsilon' \simeq \gamma(1-\cos\theta_O)d\epsilon_O'$.
The Compton scattering differential rate in the frame $_O$ is
\begin{equation}
\frac{dN}{dt_O d\epsilon_Od\Omega_O d\epsilon_O'} =c \frac{d\sigma(\epsilon_O)}{d\Omega_O}\delta(\epsilon_O'-\epsilon_O) \frac{dn_O}{d\epsilon_O}, \label{eq: diff rate O}
\end{equation}
Here the $d\sigma(\epsilon_O)/d\Omega_O$ is the Klein-Nishina cross section.
%where $dn_O/d\epsilon_O$ is the differential photon number density in the frame $_O$.
The photon density spectrum $dn_O/d\epsilon_O$ in the $_O$ frame can be related with the isotropic differential photon density in the laboratory frame by the Lorentz invariance of the ratio $dn/\epsilon$.
\begin{equation}
\frac{1}{\epsilon_O}\frac{dn_O}{d\epsilon_O} = \frac{1}{\epsilon}\frac{dn}{d\epsilon_O}. \label{eq: inv den}
\end{equation}
The isotropic differential photon density in the laboratory frame reads $dn=\frac{1}{2}n(\epsilon) d\cos\phi$, where $n(\epsilon)$ is the density of photons of a given energy $\epsilon$.
According to Eq. (\ref{eq: ene to O}), the incident angle in the laboratory frame results in a change in the photon energy in the $_O$ frame as $\vert d\epsilon_O/d\cos\phi \vert \simeq \gamma\epsilon$.
One thus obtain from Eq. (\ref{eq: inv den})
\begin{equation}
\frac{dn_O}{d\epsilon_O} = \frac{\epsilon_O}{2\gamma \epsilon^2} n(\epsilon). \label{eq: den to O}
\end{equation}
By combining Eqs. (\ref{eq: den to O}) and (\ref{eq: diff rate O}) with Eq. (\ref{eq: diff rate lab}), the Compton scattering differential rate reads~\citep{Blumenthal_RMP_1970}

\begin{equation}
\frac{dN}{dtd\mathcal{E}'} = \frac{3\sigma_T c}{4\gamma}\frac{n(\epsilon)}{\epsilon} f(\Gamma, \mathcal{E}'),
\end{equation}
where $\mathcal{E}'=\epsilon'/\epsilon'_{\mathrm{max}}$ is the scattered photon energy normalised to its maximum $\epsilon'_{\mathrm{max}}=\gamma\Gamma/(1+\Gamma)$.
The parameter $\Gamma=4\epsilon\gamma$ relates to the energy of the scattering photons in the electron rest frame and distinguishes two regimes: i) Thomson limit $\Gamma\ll1$ and ii) extreme Klein-Nishina limit $\Gamma\gg1$.
The scattered photon distribution function reads
\begin{equation}
f(\Gamma,\mathcal{E}') = 2q\log q +(1+2q)(1-q)+\frac{1}{2}\frac{\Gamma^2 q^2}{1+\Gamma q}(1-q)
\end{equation}
where $q=\mathcal{E}'/[1+\Gamma(1-\mathcal{E}')]$.

\section{Relaxation to thermal equilibrium of a photon gas: Kompaneets equation} \label{App: Kompaneets}

We recall the main steps in the derivation of Eq. (\ref{eq: linKomp}), see Ref. \citep{Kompaneets_JETP_1957}.
In the Thomson limit $\hbar\omega\ll mc^2$, the energy exchange of one transition is small compared to the energy of the photon $\delta\omega=\vert\omega'-\omega\vert\ll\omega$.
The energy exchange over one Compton event is
\begin{eqnarray}
\hbar\delta\omega &=& \hbar\omega\frac{c{\bf p}\cdot(\hat{\bf k}'-\hat{\bf k})-\hbar\omega(1-\hat{\bf k}'\cdot\hat{\bf k})}{\gamma mc^2+\hbar\omega(1-\hat{\bf k}'\cdot\hat{\bf k})-c{\bf p}\cdot\hat{\bf k}} \\
&\simeq& \hbar\omega\left[\frac{\bf p}{mc}\cdot(\hat{\bf k}'-\hat{\bf k})-\frac{\hbar\omega}{mc^2}(1-\hat{\bf k}'\cdot\hat{\bf k})\right]
\end{eqnarray}
where $\hat{\bf k}={\bf k}/k$ and $\hat{\bf k}'={\bf k}'/k'$ are the unit vectors that identify the photon propagation direction before and after scattering.
In such regime, the functions $f_e'$ and $n'$ can be expanded to second order in the small parameter $\delta\omega$ allowing the reduction of the full collision operator, Eq. (\ref{eq: fullBoltz}), to a Fokker-Planck equation 
\begin{eqnarray}
\frac{\partial n}{c\partial t} &\simeq& \left[\frac{\partial n}{\partial \xi} +n(1+n)\right] \int d{\bf p} \frac{d\sigma}{d\Omega}f_e \frac{\hbar \delta\omega}{k_BT} \nonumber \\
&&  +\left[\frac{\partial^2 n}{\partial \xi^2} +(1+n)\left(2\frac{\partial n}{\partial \xi} +n\right)\right] \int d{\bf p} \frac{d\sigma}{d\Omega}f_e \left(\frac{\hbar \delta\omega}{k_BT}\right)^2, \nonumber \\
\end{eqnarray}
where the electron distribution function is assumed to be Maxwellian, and $\xi=\hbar\omega/ k_BT$ is the energy of the photon normalised to the electron temperature.
The expansion parameter $\delta\omega$ is small in the laboratory frame only if it is also small in the proper frame of each electron.
This holds for non relativistic electron temperatures $k_BT\ll mc^2$.
The two integrals in $\delta\omega$ and in $\delta\omega^2$ can be evaluated assuming the differential cross section in the Thomson limit
\begin{equation}
\frac{d\sigma}{d\Omega} = \frac{r_e^2}{2}\left( 1+ \cos^2\theta\right).
\end{equation}
Then, the time evolution of the average occupation photon number $n$ reads 
\begin{equation}
\xi^2\frac{\partial n}{\partial y} = \frac{\partial}{\partial \xi} \left[\xi^4\left(\frac{\partial n}{\partial \xi} +n +n^2\right)\right].
\end{equation}
where $y=t/t_C$ is the time normalised to $t_C=mc/\sigma_Tn_ek_BT$, $n_e$ is the electron gas density.
The time $t_C$ is the characteristic relaxation time of the process and the thermal equilibrium is reached when $y > 1$.

In regimes where the photon occupation number is small, $n\ll1$, the photon electron gas interaction is mediated by single Compton scattering events and the equation reduces to its linear form
\begin{equation}
\xi^2\frac{\partial n}{\partial y} = \frac{\partial}{\partial \xi} \left[\xi^4\left(\frac{\partial n}{\partial \xi} +n \right)\right].
\end{equation}
In terms of the photon energy distribution function $f=\xi^2 n$, the linear Kompaneets equation reads
\begin{equation}
\frac{\partial f}{\partial y} = \frac{\partial }{\partial \xi} \left[\xi^2 \frac{\partial f}{\partial \xi} +\left(\xi^2-2\xi\right) f \right]
\end{equation}

\bibliographystyle{jpp}

\bibliography{CMP_mod}

\end{document}